\documentclass[12pt,twoside]{article}

\usepackage{pslatex}	
\usepackage{amsfonts}
\usepackage{amsmath}
\usepackage{amsthm}
\usepackage{amscd}
\usepackage{cite}
\usepackage{epsf}
\usepackage{epsfig}
\usepackage{a4}

\def\beq{\begin{equation}}
\def\eeq{\end{equation}}
\def\bea{\begin{eqnarray}}
\def\eea{\end{eqnarray}}

\def\beann{\begin{eqnarray*}}
\def\eeann{\end{eqnarray*}}

\let\a=\alpha \let\be=\beta \let\g=\gamma \let\de=\delta
  \let\h=\eta 

\let\dh=\vartheta \let\k=\kappa \let\la=\lambda \let\m=\mu
\let\n=\nu \let\x=\xi \let\p=\pi \let\r=\rho \let\s=\sigma
 
  \let\PH=\Phi \let\Ps=\Psi
  
\let\La=\Lambda  \let\D=\Delta

\let\qd=\quad  

\def\epp{\, .}
\def\epc{\, ,}

\def\tst#1{{\textstyle #1}}

\theoremstyle{plain}
\newtheorem{theorem}{Theorem}

\newtheorem{proposition}{Proposition}
\newtheorem{corollary}{Corollary}
\newtheorem*{corollary*}{Corollary}

\theoremstyle{definition}

\newtheorem*{example}{Example}

\def\2{\frac{1}{2}} \def\4{\frac{1}{4}}

\def\6{\partial}

\def\+{\dagger}

\def\<{\langle} \def\>{\rangle}

  \def\CT{{\cal T}}

\def\i{{\rm i}}

\def\re{{\rm e}}

\DeclareMathOperator{\sh}{sh}
\DeclareMathOperator{\ch}{ch}
\DeclareMathOperator{\tgh}{th}

\DeclareMathOperator{\cth}{cth}

\DeclareMathOperator{\tr}{tr}

\DeclareMathOperator{\End}{End}
\DeclareMathOperator{\id}{id}

\def\detq{{\rm det}_q}

\renewcommand{\tilde}{\widetilde}

\begin{document}

\thispagestyle{empty}

\begin{center}

{\Large {\bf Surface free energy for systems with integrable boundary
conditions\\}}

\vspace{7mm}

{\large
Frank G\"{o}hmann$^1$%
\footnote[2]{e-mail: goehmann@physik.uni-wuppertal.de},
Michael Bortz$^2$\footnote[1]{e-mail: michael.bortz@anu.edu.au} and
Holger Frahm$^3$\footnote[3]{e-mail: frahm@itp.uni-hannover.de}
\\}

\vspace{5mm}

$^1$Fachbereich C -- Physik, Bergische Universit\"at Wuppertal,\\
42097 Wuppertal, Germany\\
$^2$Department of Theoretical Physics, Australian National University,\\
Canberra ACT 0200, Australia\\
$^3$Institut f\"ur Theoretische Physik, Universit\"at Hannover,\\
30167 Hannover, Germany\\

\vspace{20mm}

{\large {\bf Abstract}}

\end{center}

\begin{list}{}{\addtolength{\rightmargin}{10mm}
               \addtolength{\topsep}{-5mm}}
\item
The surface free energy is the difference between the free
energies for a system with open boundary conditions and the same
system with periodic boundary conditions. We use the quantum transfer
matrix formalism to express the surface free energy in the
thermodynamic limit of systems with integrable boundary conditions
as a matrix element of certain projection operators. Specializing to
the XXZ spin 1/2 chain we introduce a novel `finite temperature
boundary operator' which characterizes the thermodynamical properties
of surfaces related to integrable boundary conditions.
\\[2ex]
\end{list}

\clearpage

\section{Introduction}
The embedding of non-magnetic impurities into spin chains has
profound effects on the low-energy properties of such systems.
Quantities such as the magnetic susceptibility will acquire corrections
of order ${\cal O}(L^0)$ which are directly accessible to experimental
observation ($L$ being the length of the chains). The dependence of
these surface contributions on temperature and magnetic field has been
a focus of theoretical work for some time now. Apart from field
theoretical approaches \cite{FuEg04,FuHi04} the exact solution of the
XXZ spin-$1/2$ chain with open boundary conditions \cite{Sklyanin88,%
ABBBQ87} has been used to address this problem \cite{SaTs95,FrZv97b,%
ZvMa04}. At zero temperature, the calculation of surface contributions
to the ground state energy and the magnetic susceptibilities by means
of the Bethe Ansatz is well established \cite{AsSu96,HQB87,BoSi05,%
FrZv97b,ZvMa04}. For finite $T$, bulk thermal properties are known from
the thermodynamic Bethe Ansatz (TBA) based on the so-called string
hypothesis \cite{Takahashi99}. This approach has been extended to
compute surface corrections to these quantities \cite{SaTs95,FrZv97b,%
ZvMa04}. The temperature dependence of the boundary susceptibility
obtained this way, however, disagrees with the field theoretical
results \cite{FuEg04,FuHi04} which, when combined with the leading
$T = 0$ contribution from the Bethe Ansatz, have been shown to provide
the correct low temperature asymptotics \cite{BoSi05}. In addition,
this approach fails to reproduce the correct high-temperature
behaviour. Since the Bethe Ansatz yields reliable results for the open
chains at $T=0$, the origin of these problems is likely to be the
improper treatment of the combinatoric entropy which is a central
quantity in the TBA. First attempts to extend the TBA for the
calculation of the surface contributions to the free energy exist
\cite{Woynarovich04}. It is still an open issue, though, whether this
approach can resolve the problems mentioned above.

In this paper we choose a different approach to obtain the surface
contribution to the free energy from the exact solution by combining
Sklyanin's work \cite{Sklyanin88} on the algebraic construction of
integrable open boundary conditions with the quantum transfer matrix
approach \cite{Suzuki85,SuIn87,Kluemper92,Kluemper93} to the
thermodynamics of solvable quantum systems in one dimension.

In section 2 we recall the results of Sklyanin's work \cite{Sklyanin88}
as far as they are needed for our purposes. Section 3 contains a short
review of the quantum transfer matrix approach to the thermodynamics
of solvable quantum lattice models. In spirit and notation we follow
\cite{GKS04a}, where more details can be found. Section 4 contains
our first expression for the surface free energy of a system connected
to Sklyanin's reflection algebra and treatable by the quantum transfer
matrix method. The formula is exposed in proposition \ref{prop:locmat}
and shows how to obtain the surface free energy as a Trotter limit
of an expectation value of a product of certain projection operators
which are local in the space where the quantum transfer matrix acts. The
expectation value is taken in a certain so-called dominant eigenstate of
the quantum transfer matrix. A second expression for the surface free
energy as an expectation value of a non-local operator in the dominant
state is derived in section 5 (see equation (\ref{surffree2})).
Again a Trotter limit must be taken, and a novel `finite temperature
boundary operator' appears. From its properties, worked out in the
remainder of section 5, we shall see that it is a generic object in
the context of the quantum transfer matrix formalism applied to open
boundary systems. From section 5 on we restrict ourselves to the
example of the XXZ chain with local spins $\2$ and only comment on
the generalization of the results later in the concluding section~6.

\section{Integrable boundary conditions}
Our construction of the finite temperature boundary operator is
based on a combination of the quantum transfer matrix approach to the
thermodynamics of `Yang-Baxter integrable' systems, as reviewed in the
following section, with Sklyanin's work \cite{Sklyanin88} on the
construction of integrable boundary conditions. Sklyanin's construction
is valid for a fairly general class of integrable systems characterized
by an $R$-matrix of the form $R(\la,\m) = R(\la - \m) \in
\End (V \otimes V)$ (V vector space with $\dim V = d \in {\mathbb N})$
which not only satisfies the Yang-Baxter equation
\begin{equation} \label{ybe}
     R_{12} (\la, \m) R_{13} (\la, \n) R_{23} (\m, \n) =
     R_{23} (\m, \n) R_{13} (\la, \n) R_{12} (\la, \m)
\end{equation}
but several additional conditions. Namely, $R(\la)$ is symmetric in
the sense that
\begin{equation} \label{lrsym}
     P R(\la) P = R(\la) \epc
\end{equation}
where $P$ is the transposition map on $V \otimes V$ ($P \, x \otimes y
= y \otimes x$). Furthermore, $R(\la)$ is unitary,
\begin{equation} \label{uni}
     R(\la) R(-\la) = \r (\la)
\end{equation}
for some complex function $\r (\la)$, and crossing unitary,
\begin{equation} \label{crossuni}
     R^{t_1} (\la) R^{t_1} (- \la - 2 \h) = \tilde \r (\la)
\end{equation}
for another complex function $\tilde \r (\la)$ and $\h$ a parameter
characterizing the $R$-matrix. The superscript $t_1$ in (\ref{crossuni})
denotes the transposition with respect to the first space in the
tensor product $V \otimes V$. Similarly, $t_2$ will denote the
transposition with respect to the second space. Then $R(\la)$ is a
symmetric matrix if
\begin{equation} \label{sym}
     R^{t_1 t_2} (\la) = R(\la) \qd \Leftrightarrow \qd
     R^{t_1} (\la) = R^{t_2} (\la) \epc
\end{equation}
which is also assumed by Sklyanin.

\begin{example}
Our chief example will be the $R$-matrix
\begin{equation} \label{rxxz}
     R(\la) = \begin{pmatrix}
                 1 & 0 & 0 & 0 \\
		 0 & b(\la) & c(\la) & 0 \\
		 0 & c(\la) & b(\la) & 0 \\
		 0 & 0 & 0 & 1
              \end{pmatrix}
\end{equation}
with
\begin{equation} \label{defbc}
     b(\la) = \frac{\sh(\la)}{\sh(\la + \h)} \epc \qd
     c(\la) = \frac{\sh(\h)}{\sh(\la + \h)} \epp
\end{equation}
This is the well-known 6-vertex model solution of the Yang-Baxter
equation (\ref{ybe}). It generates the Hamiltonian of the
XXZ chain with local spins $\2$.  $R(\la)$ satisfies
(\ref{lrsym})-(\ref{sym}) with
\begin{equation} \label{rhorho}
     \r (\la) = 1 \epc \qd
     \tilde \r (\la) = \frac{\sh(\la) \sh(\la + 2 \h)}
                            {\sh^2 (\la + \h)} \epp
\end{equation}
\end{example}

Every solution $R(\la,\m)$ of the Yang-Baxter equation (\ref{ybe})
determines the structure of a Yang-Baxter algebra with generators
$T^\a_\be (\la)$, $\a, \be = 1, \dots, d$, through the relations
\begin{equation} \label{yba}
     R_{12} (\la,\m) T_1 (\la) T_2(\m)
        = T_2 (\m) T_1 (\la) R_{12} (\la,\m) \epc
\end{equation}
where $T_1 (\la) = T(\la) \otimes \id_V$, $T_2 (\la) =
\id_V \otimes T(\la)$. The matrix $T(\la)$ of generators is called
the monodromy matrix. The map defined by
\begin{equation} \label{antipode}
     T^a (\la) = (T^{-1})^t (\la)
\end{equation}
is an algebra automorphism, and Sklyanin calls it the antipode. He
restricts his considerations to representations of the Yang-Baxter
algebra (\ref{yba}) possessing the crossing symmetry
\begin{equation} \label{crossing}
     (T^a)^a (\la) = \dh (\la) T(\la - 2 \h) \epp
\end{equation}
Here $\dh(\la)$ is a complex function depending on the representation
and $\h$ is the same parameter as in (\ref{crossuni}).

Sklyanin's construction of integrable systems with boundaries is based
on the representations of two `reflection algebras' $\CT^{(-)}$ and
$\CT^{(+)}$ which, for given $R(\la)$ satisfying (\ref{ybe})-%
(\ref{sym}), are defined by the relations
\begin{subequations}
\label{refl}
\begin{align} \label{lrefl}
     R_{12} & (\la - \m) \CT_1^{(-)} (\la) R_{12} (\la + \m)
        \CT_2^{(-)} (\m) \notag \\ & \mspace{117.mu}
	= \CT_2^{(-)} (\m) R_{12} (\la + \m)
	  \CT_1^{(-)} (\la) R_{12} (\la - \m) \epc \\[1ex]
     R_{12} & (- \la + \m) \CT_1^{(+) \, t_1} (\la)
        R_{12} (- \la - \m - 2\h) \CT_2^{(+) \, t_2} (\m) \notag
        \\ \label{rrefl} & \mspace{117.mu}
	= \CT_2^{(+) \, t_2} (\m) R_{12} (- \la - \m - 2\h)
	  \CT_1^{(+) \, t_1} (\la) R_{12} (- \la + \m) \epp
\end{align}
\end{subequations}
We shall call $\CT^{(-)}$ and $\CT^{(+)}$ left and right reflection
algebras, respectively. $\CT^{(-)}$ and $\CT^{(+)}$ are isomorphic
and several isomorphisms are explicitly known \cite{Sklyanin88}.
Using (\ref{refl}) and (\ref{lrsym})-(\ref{sym}) Sklyanin proved
the basic
\begin{theorem} \label{theo:commute}
The functions (`transfer matrices')
\begin{equation}
     t(\la) = \tr \CT^{(+)} (\la) \CT^{(-)} (\la)
\end{equation}
defined in the tensor product $\CT^{(+)} \otimes \CT^{(-)}$ form
a commutative family, i.e.
\begin{equation}
     [t(\la), t(\m)] = 0 \qd \text{for all $\la, \m \in {\mathbb C}$}
        \epp
\end{equation}
\end{theorem}

Thus, given a pair of representations of $\CT^{(+)}$ and $\CT^{(-)}$
the matrix $t(\la)$ provides a generating function of `quantum
integrals of motion'. Further following Sklyanin \cite{Sklyanin88} we
shall now review a set of basic results on the representation theory
of $\CT^{(+)}$ and $\CT^{(-)}$ that will be needed in the sequel.
\begin{proposition} \label{prop:comult}
Let $\tilde \CT^{(\pm)} (\la)$ two representations of $\CT^{(\pm)}$,
respectively, in spaces $\tilde W^\pm$, and $T^{(\pm)} (\la)$ two
representations of the Yang-Baxter algebra (\ref{yba}) in $W^\pm$
such that $T^{(+)} (\la)$ satisfies (\ref{crossing}). Then
\begin{subequations}
\label{reppm}
\begin{align}
     \CT^{(-)} (\la) & = T^{(-)} (\la) \tilde \CT^{(-)} (\la)
                         T^{(-) - 1} (- \la) \epc \\ \label{repplus}
     \CT^{(+) t} (\la) & = T^{(+) \, t} (\la)
                           \tilde \CT^{(+) \, t} (\la)
                           T^{(+) \, a} (- \la)
\end{align}
\end{subequations}
are representations of $\CT^{(\pm)}$ in $\tilde W^\pm \otimes W^\pm$.
\end{proposition}
Proposition \ref{prop:comult} allows one to construct integrable open
boundary conditions for known models with $L$-matrix $L(\la)$.
These are connected with the simplest representations, namely
c-number matrices, of $\CT^{(\pm)}$. Given two representations
$K^{(\pm)} (\la)$ of $\CT^{(\pm)}$ in ${\mathbb C}$ and an $L$-matrix
representation $L(\la)$ (acting on `a local quantum space') of the
Yang-Baxter algebra (\ref{yba}), define
\begin{align} \label{tm}
     T^{(-)} (\la) & = L_M (\la) \dots L_1 (\la) \epc \\ \label{tp}
     T^{(+)} (\la) & = L_L (\la) \dots L_{M+1} (\la) \epc \\
     \tilde \CT^{(\pm)} (\la) & = K^{(\pm)} (\la) \epp \label{tpmkpm}
\end{align}
Then, by proposition \ref{prop:comult}, $\CT^{(-)} (\la) =
T^{(-)} (\la) \tilde \CT^{(-)} (\la) T^{(-) - 1} (- \la)$ and
$\CT^{(+) t} (\la) =$ \linebreak  $ T^{(+) \, t} (\la)
\tilde \CT^{(+) t} (\la) T^{(+) \, a} (- \la)$ are representations of
$\CT^{(\pm)}$ and, according to theorem~\ref{theo:commute}, $t (\la)
= \tr \CT^{(+)} (\la) \CT^{(-)} (\la)$ generates a commutative family
of operators. Moreover, the following proposition holds.
\begin{proposition} \label{prop:genf}
The generating function $t (\la) = \tr \CT^{(+)} (\la) \CT^{(-)} (\la)$
with $\CT^{(\pm)} (\la)$ defined by (\ref{reppm})-(\ref{tpmkpm}) can
be written as
\begin{equation} \label{deft}
     t(\la) = \tr K^{(+)} (\la) T(\la) K^{(-)} (\la) T^{-1} (- \la)
              \epc
\end{equation}
where $T(\la) = T^{(+)} (\la) T^{(-)} (\la) = L_L (\la) \dots
L_1 (\la)$, and is thus independent of the factorization (\ref{tm}),
(\ref{tp}) of $T(\la)$ into $T^{(+)} (\la)$ and $T^{(-)} (\la)$.
\end{proposition}

Let us now consider any fundamental model associated with a regular
$R$-matrix satisfying $R(0) = P$ (for a pedagogical review of
such type of models see e.g.\ chapter 12 of \cite{thebook}). Its
$L$-matrices have matrix elements ${L_j}^\a_\be (\la) =
R^{\a \g}_{\be \de} (\la) {e_j}_\g^\de$, $j = 1, \dots L$, where the
$e_\a^\be$ are matrices with a single non-zero entry 1 at the
intersection of the $\a$th row and the $\be$th column, $\a, \be =
1, \dots d$. The Hamiltonian associated with a fundamental model is
\begin{align} \label{hper}
     H & = (\tr T(0))^{-1} \6_\la \tr T(\la) \bigr|_{\la = 0}
         = \sum_{j=1}^{L - 1} H_{j j+1} + H_{L 1} \epc \\ \label{hloc}
     H_{jk} & = \6_\la (P R)_{jk} (\la) \bigr|_{\la = 0} \epp
\end{align}
The occurrence of the term $H_{L1}$ on the right hand side means that
we are dealing with periodic boundary conditions. Using $t(\la)$
defined in (\ref{deft}) instead of the `row-to-row transfer matrix'
$\tr T(\la)$ as a generating function, we obtain the same Hamiltonian
but with different boundary terms.
\begin{proposition}
Consider the generating function $t(\la)$ defined in proposition
\ref{prop:genf} with $T(\la)$ a monodromy matrix of a fundamental
model and $K^{(\pm)} (\la)$ c-number representations of the right
and left reflection algebras (\ref{refl}) satisfying
\begin{equation} \label{norm}
     \tr K^{(+)} (\la) = 1 \epc \qd K^{(-)} (0) = \id_V \epp
\end{equation}
Then
\begin{equation} \label{tasymp}
     t(\la) = 1 + 2 \la {\cal H} + {\cal O} (\la^2) \epc
\end{equation}
where
\begin{equation} \label{hopen}
     {\cal H} = \sum_{j=1}^{L - 1} H_{j j+1}
         + \frac{\6_\la K_1^{(-)} (\la)|_{\la = 0}}{2}
	 + \tr_0 K_0^{(+)} (0) H_{L 0}
\end{equation}
and $H_{jk}$ is the same as in (\ref{hloc}).
\end{proposition}
Here the two on-site terms on the right hand side of (\ref{hopen})
may be interpreted as external fields acting at the boundary sites
1 and $L$ of the chain. In particular, it may be possible that both
of these terms vanish for certain choices of $K^{(\pm)} (\la)$ (see
next example below). Then we are dealing with a chain which is cut
at the link between sites $L$ and 1, and hence has reflecting ends.
Note that, unlike Sklyanin, we have assumed that $\tr K^{(+)} (\la)
= 1$ for all $\la \in {\mathbb C}$. This is not much of a restriction
since the Hamiltonian turns out to be trivial if $\tr K^{(+)} (\la)$
vanishes identically, and otherwise the trace can be replaced by 1,
since the reflection equations (\ref{refl}) are homogeneous. With
$\tr K^{(+)} (\la) = 1$ the expression (\ref{hopen}) for the
Hamiltonian looks slightly simpler, and we have $t(0) = 1$ which will
become important later, when we proceed to the quantum transfer matrix
formalism.

\begin{example}
Diagonal solution of the reflection equations for the XXZ chain
\cite{Sklyanin88}. Let
\begin{equation} \label{kmother}
     K(\la,\x) = I_2 + \tgh (\la) \cth (\x) \s^z \epc
\end{equation}
where $I_2$ denotes the $2 \times 2$ unit matrix and $\s^z = e_1^1
- e_2^2$ the diagonal Pauli matrix. Then, up to the multiplication
with an arbitrary function,
\begin{equation} \label{kxxz}
     K^{(-)} (\la) = K(\la, \x^-) \epc \qd
     K^{(+)} (\la) = \tst{\2} K(\la + \h, \x^+)
\end{equation}
are the unique diagonal c-number solutions of (\ref{refl}) with
$R$-matrix (\ref{rxxz}). They are normalized such as to fulfill
(\ref{norm}). Inserting (\ref{rxxz}), (\ref{defbc}), (\ref{hloc}) and
(\ref{kmother}), (\ref{kxxz}) into (\ref{hopen}) we obtain
\begin{multline}
     2 \sh(\h) {\cal H} = \sum_{j=1}^{L-1}
                   \bigl[ \s_j^x \s_{j+1}^x + \s_j^y \s_{j+1}^y
		      + \ch(\h) (\s_j^z \s_{j+1}^z - 1) \bigr] \\
		      + \sh(\h) \bigl[ \cth(\x^-) \s_1^z
		                       + \cth(\x^+) \s_L^z \bigr]
		      - \ch(\h) \epp
\end{multline}
This is the open XXZ spin-$\2$ chain with anisotropy parameter $\D =
\ch(\h)$ and longitudinal local magnetic fields $h^\mp = \sh(\h)
\cth(\x^\mp)$ acting at the left and right boundaries. Note that
the boundary fields vanish for $\x^\pm = \i \p/2$.
\end{example}

\section{Quantum transfer matrix and free energy}
We recall that every solution $R(\la,\m) = R(\la - \m)$ of the
Yang-Baxter equation (\ref{ybe}) defines an integrable vertex model
with row-to-row transfer matrix
\begin{equation} \label{trtr}
     \tr T(\la) = \tr_0 R_{0L} (\la,0) \dots R_{01} (\la,0)
                = \tr L_L (\la) \dots L_1 (\la) \epp
\end{equation}
If $R(\la)$ is regular, then $\tr T(\la)$ generates a local Hamiltonian
with periodic boundary conditions via (\ref{hper}), (\ref{hloc}).
Hence, knowing the eigenvalues of $\tr T(\la)$ means to know the
eigenvalues of the Hamiltonian (\ref{hper}).

The row-to-row transfer matrix (\ref{trtr}) is not the most convenient
tool, however, if one wishes to study the thermodynamic properties of
the Hamiltonian (\ref{hper}), since, for the calculation of the
partition function, all eigenvalues of $H$, and hence all eigenvalues
of $\tr T(\la)$, are needed. Fortunately, another auxiliary vertex
model can be constructed by means of the same $R$-matrix, whose
partition function in a certain so-called Trotter limit is equal to the
partition function of the Hamiltonian (\ref{hper}) and, asymptotically
for large $L$, is determined by a single leading eigenvalue of the
associated so-called quantum transfer matrix \cite{Suzuki85,SuIn87}.

For $j = 1, \dots, L$ and $\be \in {\mathbb C}$ we define the
monodromy matrices
\begin{equation} \label{monoqtm}
     T^{QTM}_j (\la ) =
	R_{\bar N j}^{t_1}
	   \bigl(- \tst{\frac{\be}{N}}, \la \bigr)
        R_{j \overline{N-1}} \bigl(\la, \tst{\frac{\be}{N}} \bigr) \dots
	R_{\bar 2 j}^{t_1} \bigl(- \tst{\frac{\be}{N}}, \la \bigr)
        R_{j \bar 1} \bigl(\la, \tst{\frac{\be}{N}} \bigr) \epp
\end{equation}
By slight abuse of the usual terminology we shall call them the
`quantum monodromy matrices'. They define representations of the
Yang-Baxter algebra as is obvious from the defining relations
(\ref{yba}) and their transpose with respect to space 1 (see
\cite{GKS04a} for details). It is easy to see (e.g.\ \cite{GKS04a})
that
\begin{equation} \label{statop}
     \lim_{N \rightarrow \infty} \tr_{\bar 1 \dots \bar N}
        T^{QTM}_1 (0) \dots T^{QTM}_L (0) = \re^{- \be H} \epc
\end{equation}
which is the statistical operator associated with the Hamiltonian
(\ref{hper}) if we identify $\be$ as the inverse temperature $1/T$. It
follows that the partition function of the $L$-site chain with
periodic boundary conditions is
\begin{equation} \label{periodicpart}
     Z_L = \lim_{N \rightarrow \infty} \tr_{\bar 1 \dots \bar N}
           \bigl( \tr T^{QTM} (0) \bigr)^L
	 = \sum_{n=0}^\infty \La_n^L (0) \epc
\end{equation}
where $\La_n (\la)$ is the $n$th eigenvalue of the quantum transfer
matrix $\tr T^{QTM} (\la)$ in the Trotter limit $N \rightarrow \infty$.
By definition $\La_0 (\la)$ is the eigenvalue with largest modulus at
$\la = 0$. It must be finitely degenerate in a vicinity of zero. For
simplicity we shall assume it to be non-degenerate. In any case, the
free energy per lattice site in the thermodynamic limit turns out to
be \cite{Suzuki85,SuIn87}
\begin{equation} \label{free}
     f = - T \ln \La_0 (0) \epp
\end{equation}

For many models, e.g.\ for the XXZ chain determined by the $R$-matrix
(\ref{rxxz}), the leading eigenvalue of the quantum transfer matrix
can be calculated for every finite Trotter number $N$ by means of
the algebraic Bethe Ansatz. The corresponding solutions of the
Bethe Ansatz equations determine auxiliary functions which can be
shown to uniquely solve certain non-linear integral equations
\cite{Kluemper92,Kluemper93,thebook} containing the Trotter
number $N$ as a mere parameter. In these equations the Trotter limit
is easily performed analytically, and an integral formula involving the
auxiliary functions can be derived for the free energy per lattice
site (\ref{free}). Furthermore, it was shown in \cite{GKS04a} that the
eigenvector $|\Ps_0\>$ corresponding to the leading eigenvalue
$\La_0 (\la)$ of the quantum transfer matrix encodes the complete
information about the state of thermal equilibrium and hence about all
thermal correlation functions. This led to the discovery of integral
representations of correlation functions of the XXZ chain at finite
temperatures \cite{GKS04a,GKS05}. We shall refer to $|\Ps_0\>$ as the
`dominant state' of the quantum transfer matrix.

\section{The surface free energy as a matrix element}
The derivation of the formula (\ref{statop}) for the statistical
operator in the previous section relies on a simple generalization
of the Euler formula for the exponential function. Consider a
sequence of operators $(X_N)_{N=1}^\infty$ on a finite dimensional
vector space that converges (in some appropriate norm) to a limit $X$.
Then
\begin{equation} \label{trotter}
     \lim_{N \rightarrow \infty} \Bigl( 1 + \frac{X_N}{N} \Bigr)^N
        = \re^X \epp
\end{equation}
We shall call this the Trotter formula.

The Trotter formula can be combined with (\ref{tasymp}) in order
to obtain the statistical operator for the Hamiltonian ${\cal H}$
of the open boundary system as a limit. Setting
\begin{equation}
     X_N = N \Bigl[ t \bigl( - \tst{\frac{\be}{2N}} \bigr) - 1 \Bigr]
\end{equation}
and using (\ref{tasymp}) we find that $\lim_{N \rightarrow \infty} X_N
= - \be {\cal H}$ and therefore, using (\ref{trotter}),
\begin{equation} \label{trotterappl}
     \lim_{\substack{N \rightarrow \infty\\ N \in 2{\mathbb N}}}
        t^\frac{N}{2} \bigl( - \tst{\frac{\be}{N}} \bigr)
	= \re^{- \be {\cal H}} \epp
\end{equation}
This observation will serve us as a starting point to express the
surface free energy as a matrix element of a product of local
operators.

For a pair of c-number solutions $K^{(\pm)} (\la)$ of the reflection
equations (\ref{refl}) let us define an operator
\begin{equation} \label{defpi}
     \Pi_{\bar k \bar \ell} (\la) = K_{\bar k}^{(+)} (\la)
        P^{\, t_1}_{\bar k \bar \ell} \, K_{\bar k}^{(-)} (\la) \epc
	\qd k, \ell = 1, \dots, N \epp
\end{equation}
\begin{proposition} \label{prop:locmat}
The surface free energy, i.e.\ the difference between the free energies
of the open system with Hamiltonian (\ref{hopen}) and the corresponding
periodically closed system with Hamiltonian (\ref{hper}) is given by
\begin{equation} \label{surffree1}
     \D F = \lim_{\substack{N \rightarrow \infty\\ N \in 2{\mathbb N}}}
             - T \ln \<\Ps_0| \Pi_{\bar 1 \bar 2}
	                      \bigl( - \tst{\frac{\be}{N}} \bigr) \dots
                              \Pi_{\overline{N-1} \bar N}
	                      \bigl( - \tst{\frac{\be}{N}} \bigr)
			      |\Ps_0\> \epc
\end{equation}
where $|\Ps_0\>$ is the dominant state of the quantum transfer matrix
$\tr T^{QTM} (\la)$ (see (\ref{monoqtm})).
\end{proposition}
\begin{proof}
Let us define
\begin{subequations}
\begin{align}
     T_{\bar k} (\la) & = R_{\bar k L} (\la,0) \dots
                          R_{\bar k 1} (\la, 0) \epc \\
     \overline{T}_{\bar k} (\la) & = R_{1 \bar k} (0,\la) \dots
                                     R_{L \bar k} (0,\la)
\end{align}
\end{subequations}
for $k = 1, \dots, N$ (remark: then $\overline{T}_{\bar k} (\la) =
T^{-1}_{\bar k} (\la)$ for unitary $R$ matrices with $\r(\la) = 1$%
\footnote{Henceforth we simply assume that $\r(\la) = 1$. This can
usually be achieved by properly normalizing the $R$-matrix and is
certainly true for the $R$-matrix (\ref{rxxz}) of the XXZ chain (see
equation (\ref{rhorho})).}). It
follows that
\begin{align}
     t(\la) & = \tr_{\bar 1} K^{(+)}_{\bar 1} (\la) T_{\bar 1} (\la)
                K^{(-)}_{\bar 1} (\la) \overline{T}_{\bar 1} (- \la)
	        \notag \\
            & = \tr_{\bar 1 \bar 2} K^{(+)}_{\bar 1} (\la)
	        P_{\bar 1 \bar 2} T_{\bar 1} (\la)
                K^{(-)}_{\bar 1} (\la) \overline{T}_{\bar 1} (- \la)
	        \notag \\
            & = \tr_{\bar 1 \bar 2} K^{(+)}_{\bar 1} (\la)
		\bigl( T_{\bar 2} (\la) P_{\bar 1 \bar 2}
		       \bigr)^{\, t_{\bar 2}}
                K^{(-)}_{\bar 1} (\la) \overline{T}_{\bar 1} (- \la)
	        \notag \\
            & = \tr_{\bar 1 \bar 2} K^{(+)}_{\bar 1} (\la)
		P_{\bar 1 \bar 2}^{t_{\bar 1}} \, K^{(-)}_{\bar 1} (\la)
		T^t_{\bar 2} (\la) \overline{T}_{\bar 1} (- \la)
	        \notag \\
            & = \tr_{\bar 1 \bar 2} \Pi_{\bar 1 \bar 2} (\la)
		T^t_{\bar 2} (\la) \overline{T}_{\bar 1} (- \la) \epp
\end{align}
Hence,
\begin{align}
        t^\frac{N}{2} \bigl( - \tst{\frac{\be}{N}} \bigr) & =
	   \tr_{\bar 1 \dots \bar N}
	   \Pi_{\overline{N-1} \bar N} \,
	   T^t_{\bar N} \bigl( - \tst{\frac{\be}{N}} \bigr)
	   \overline{T}_{\overline{N-1}}
	   \bigl( \tst{\frac{\be}{N}} \bigr) \dots
	   \Pi_{\bar 1 \bar 2}
	   T^t_{\bar 2} \bigl( - \tst{\frac{\be}{N}} \bigr)
	   \overline{T}_{\bar 1}
	   \bigl( \tst{\frac{\be}{N}} \bigr) \notag \\
	   & = 
	   \tr_{\bar 1 \dots \bar N}
	   \Pi_{\bar 1 \bar 2} \Pi_{\bar 3 \bar 4} \dots
	   \Pi_{\overline{N-1} \bar N} \,
	   T^t_{\bar N} \bigl( - \tst{\frac{\be}{N}} \bigr)
	   \overline{T}_{\overline{N-1}}
	   \bigl( \tst{\frac{\be}{N}} \bigr) \dots
	   T^t_{\bar 2} \bigl( - \tst{\frac{\be}{N}} \bigr)
	   \overline{T}_{\bar 1}
	   \bigl( \tst{\frac{\be}{N}} \bigr) \notag \\
	   & = 
	   \tr_{\bar 1 \dots \bar N}
	   \Pi_{\bar 1 \bar 2} \Pi_{\bar 3 \bar 4} \dots
	   \Pi_{\overline{N-1} \bar N}
	   T^{QTM}_1 (0) \dots T^{QTM}_L (0) \epp
\end{align}
For the last line compare equation (24) of \cite{GKS04a}. For
simplicity we left out the argument $- \be/N$ of the operators $\Pi$
here. Using (\ref{trotterappl}) we obtain the partition function
${\cal Z}_L$ of the $L$-site open system as
\begin{multline} \label{openpart}
     {\cal Z}_L =
           \lim_{\substack{N \rightarrow \infty\\ N \in 2{\mathbb N}}}
	   \tr_{\bar 1 \dots \bar N}
	   \Pi_{\bar 1 \bar 2}
	   \bigl( - \tst{\frac{\be}{N}} \bigr)
	   \dots \Pi_{\overline{N-1} \bar N}
	   \bigl( - \tst{\frac{\be}{N}} \bigr)
	   \bigl( \tr T^{QTM} (0) \bigr)^L \\
	   =
	   \lim_{\substack{N \rightarrow \infty\\ N \in 2{\mathbb N}}}
	     \sum_{n=0}^{d^N - 1} \<\Ps_n| \Pi_{\bar 1 \bar 2}
	     \bigl( - \tst{\frac{\be}{N}} \bigr) \dots
	     \Pi_{\overline{N-1} \bar N}
	     \bigl( - \tst{\frac{\be}{N}} \bigr) |\Ps_n\>
	     \La_n^L (0) \epc
\end{multline}
where $\<\Ps_n|$ and $|\Ps_n\>$ are the left and right eigenvectors
of the quantum transfer matrix. Comparing (\ref{periodicpart}) and
(\ref{openpart}) we conclude that
\begin{equation}
     \lim_{L \rightarrow \infty} \frac{{\cal Z}_L}{Z_L} =
        \<\Ps_0| \Pi_{\bar 1 \bar 2}
	         \bigl( - \tst{\frac{\be}{N}} \bigr) \dots
		 \Pi_{\overline{N-1} \bar N}
		 \bigl( - \tst{\frac{\be}{N}} \bigr) |\Ps_0\>
\end{equation}
which completes the proof of proposition \ref{prop:locmat}.
\end{proof}
Let us examine the structure of the operator $\Pi$, equation
(\ref{defpi}). Its components are
\begin{equation}
     \Pi^{\a \g}_{\be \de} (\la)
        = K^{(+)} (\la)^\a_\g K^{(-)} (\la)^\de_\be \epp
\end{equation}
This suggests to introduce row and column vectors $|-\>$ and $\<+|$
with components
\begin{equation}
     _{\a \be}|-\> = K^{(-)} (\la)^\be_\a \epc \qd
        \<+|^{\a \be} = K^{(+)} (\la)^\a_\be \epp
\end{equation}
Then
\begin{equation}
     \Pi^{\a \g}_{\be \de} (\la) = _{\be \de}|-\> \<+|^{\a \g}
        = \bigl(|-\> \otimes \<+|\bigr)^{\a \g}_{\be \de} \epp
\end{equation}
As usual we leave out the tensor product sign and write $\Pi (\la) =
|-\>\<+|$. From the definitions of $|-\>$ and $\<+|$ it follows
that $|-\>\<+|$ is proportional to a one-dimensional projector,
\begin{equation}
     \Pi^2 (\la) = \<+|-\> \, \Pi (\la) \epc \qd
        \<+|-\> = \tr K^{(+)} (\la) K^{(-)} (\la) \epp
\end{equation}

\begin{example}
Let us illustrate our result with the example of the diagonal
`$K$-matrices' (\ref{kxxz}) of the XXZ chain. For these matrices
\begin{align}
     |-\> & = \frac{1}{\sh(\x^-) \ch(\la)}
              \bigl( \sh(\x^- + \la), 0, 0, \sh(\x^- - \la) \bigr) \epc
	      \\
     \<+| & = \frac{1}{2 \sh(\x^+) \ch(\la + \h)}
              \bigl( \sh(\x^+ + \la + \h), 0, 0,
	             \sh(\x^+ - \la - \h) \bigr)^t \epp
\end{align}
Hence,
\begin{multline}
     \Pi (\la) = |-\>\<+|
               = \frac{1}{2 \sh(\x^+) \sh(\x^-) \ch(\la) \ch(\la + \h)}
	         \\[2ex]
     \begin{pmatrix}
        \sh(\x^- + \la) \sh(\x^+ + \la + \h) & 0 & 0 &
        \sh(\x^- - \la) \sh(\x^+ + \la + \h) \\
	0 & 0 & 0 & 0 \\
	0 & 0 & 0 & 0 \\
        \sh(\x^- + \la) \sh(\x^+ - \la - \h) & 0 & 0 &
        \sh(\x^- - \la) \sh(\x^+ - \la - \h)
     \end{pmatrix}
\end{multline}
and
\begin{equation}
     \<+|-\> = 1 + \cth(\x^-) \cth(\x^+) \tgh(\la) \tgh(\la + \h) \epp
\end{equation}
As we remarked above, open boundaries with vanishing boundary fields
are realized for $\x^+ = \x^- = \i \p/2$. Then $\sh(\i \p/2 + \la)
= \i \ch(\la)$ and thus
\begin{equation}
     \Pi (\la) = \2 \begin{pmatrix}
                    1 & 0 & 0 & 1 \\
		    0 & 0 & 0 & 0 \\
		    0 & 0 & 0 & 0 \\
		    1 & 0 & 0 & 1
		    \end{pmatrix} \epc
\end{equation}
which is a $\la$-independent true projection operator.
\end{example}

\section{The finite temperature boundary operator}
In proposition \ref{prop:locmat} we have expressed the surface free
energy as an expectation value of a product of certain projection
operators $\Pi$ in the dominant state of the quantum transfer matrix.
The projectors $\Pi$ are local in the space where the quantum transfer
matrix acts. Having in mind that, within the algebraic Bethe Ansatz,
the dominant state is constructed by acting with certain
creation-operator-like monodromy matrix elements on a `Fock vacuum',
we would like to express the product of projection operators in
equation (\ref{surffree1}) as well in terms of elements of the monodromy
matrix. This is possible by means of an inversion formula obtained
in \cite{KMT99a,GoKo00}.

Instead of working out all steps in full generality we shall stick
from now on with the example of the XXZ chain. The final result
will be of sufficiently general appearance, and we shall indicate
how to move on to other systems. In case of the XXZ chain the auxiliary
vector space $V$ is two-dimensional, and the quantum monodromy matrix
(\ref{monoqtm}) becomes a $2 \times 2$ matrix when expressed in
terms of the $L$-matrices
\begin{align} \label{lmatrix}
     {\cal L}_n (\la,\m) &  =
        \begin{pmatrix}
	   {e_n}_1^1 + b(\la - \m) {e_n}_2^2 & c(\la - \m) {e_n}_2^1 \\
	   c(\la - \m) {e_n}_1^2 & b(\la - \m) {e_n}_1^1 + {e_n}_2^2
        \end{pmatrix} \epc \\[2ex]
     \tilde {\cal L}_n (\m,\la) &  =
        \begin{pmatrix}
	   {e_n}_1^1 + b(\m - \la) {e_n}_2^2 & c(\m - \la) {e_n}_1^2 \\
	   c(\m - \la) {e_n}_2^1 & b(\m - \la) {e_n}_1^1 + {e_n}_2^2
        \end{pmatrix} \epc
\end{align}
namely,
\begin{equation}
     T^{QTM} (\la) =
        \tilde {\cal L}_N \bigl(- \tst{\frac{\be}{N}}, \la \bigr)
	{\cal L}_{N-1} \bigl(\la, \tst{\frac{\be}{N}} \bigr) \dots
        \tilde {\cal L}_2 \bigl(- \tst{\frac{\be}{N}}, \la \bigr)
	{\cal L}_1 \bigl(\la, \tst{\frac{\be}{N}} \bigr) \epp
\end{equation}

The aforementioned inversion formula \cite{KMT99a,GoKo00} applies
to the monodromy matrix corresponding to the usual row-to-row
transfer matrix, not to the quantum monodromy matrix. Fortunately
however, in case of the XXZ chain the different types of monodromy
matrices are simply related by means of the crossing symmetry.
\begin{equation} \label{lcrossing}
     \frac{\tilde {\cal L}_j (\m,\la)}{b(\m - \la)} =
        \s_j^y {\cal L}_j (\la, \m + \h) \s_j^y \epc
\end{equation}
where $\s_j^y$ is the Pauli matrix $\s^y = \bigl( \begin{smallmatrix}
0 & - \i \\ \i & \mspace{12.mu} 0 \end{smallmatrix} \bigr)$ acting on
site $j$. Setting 
\begin{equation}
     S = \s_N^y \s_{N-2}^y \dots \s_2^y \epc \qd
        a(\la) = b^\frac{N}{2} \bigl(- \tst{\frac{\be}{N}} - \la \bigr)
	         \epc
\end{equation}
we find that
\begin{multline} \label{tqtmrtr}
     \frac{T^{QTM} (\la)}{a(\la)} = S
        {\cal L}_N \bigl(\la, \h - \tst{\frac{\be}{N}}\bigr)
        {\cal L}_{N-1} \bigl(\la, \tst{\frac{\be}{N}}\bigr) \dots
        {\cal L}_2 \bigl(\la, \h - \tst{\frac{\be}{N}}\bigr)
        {\cal L}_1 \bigl(\la, \tst{\frac{\be}{N}}\bigr) S \\
	= S T(\la|\x_1,\dots,\x_N) S \epc
\end{multline}
where $T(\la|\x_1, \dots, \x_N) =: T(\la)$ is the inhomogeneous
row-to-row transfer matrix with inhomogeneities
\begin{equation}
     \x_j = \begin{cases}
               \h - \tst{\frac{\be}{N}} & \text{if $j$ even} \\
               \tst{\frac{\be}{N}} & \text{if $j$ odd}
            \end{cases} \epp
\end{equation}

The right hand side of (\ref{tqtmrtr}) admits the application of
the inversion formula \cite{KMT99a,GoKo00}
\begin{equation}
     {e_n}^\a_\be = \Bigl[\prod_{j=1}^{n-1} \tr T(\x_j) \Bigr]
                    T^\a_\be (\x_n)
		    \Bigl[\prod_{j=n+1}^N \tr T(\x_j) \Bigr] \epp
\end{equation}
The left hand side is the rescaled quantum monodromy matrix. We
abbreviate it as
\begin{equation} \label{rescqtm}
     \frac{T^{QTM} (\la)}{a(\la)} =: \tilde T^{QTM} (\la) 
        =: \begin{pmatrix} A(\la) & B(\la) \\ C(\la) & D(\la)
	   \end{pmatrix} \epp
\end{equation}
Taking into account that
\begin{equation}
     S \, {e_n}^\a_\be S\,  = {(\s^y)^{n+1}}^\a_\g \, {e_n}^\g_\de \,
        {(\s^y)^{n+1}}^\de_\be \epc \qd S^2 = 1
\end{equation}
we obtain the following inversion formula for the rescaled quantum
monodromy matrix (\ref{rescqtm}).
\begin{proposition}
\begin{equation}
     {e_n}^\a_\be =
        \Bigl[\prod_{j=1}^{n-1} \tr \tilde T^{QTM} (\x_j) \Bigr]
	\bigl[ (\s^y)^{n+1}
	       \tilde T^{QTM} (\x_n) (\s^y)^{n+1} \bigr]^\a_\be
        \Bigl[\prod_{j=n+1}^N \tr \tilde T^{QTM} (\x_j) \Bigr] \epp
\end{equation}
\end{proposition}
This has an immediate application to our problem of expressing the
projection operator $\Pi$, equation (\ref{defpi}), in terms of
monodromy matrix elements,
\begin{align} \label{pitau}
     \Pi_{2n-1\, 2n} \bigl(- \tst{\frac{\be}{N}}\bigr) & =
        \Bigl[\prod_{j=1}^{2n-2} \tr \tilde T^{QTM} (\x_j) \Bigr]
        \tau \bigl(- \tst{\frac{\be}{N}}\bigr)
        \Bigl[\prod_{j=2n+1}^N \tr \tilde T^{QTM} (\x_j) \Bigr] \epc
	\\[1ex] \label{deftau}
     \tau (\la) & = \tr K^{(-)} (\la) \, \tilde T^{QTM} (- \la) \,
        K^{(+)} (\la) \, \s^y \tilde{T}^{QTM \, t} (\la + \h) \s^y \epp
\end{align}
We shall call $\tau$ the `finite temperature boundary operator'.
Equation (\ref{pitau}) implies that
\begin{equation}
     \Pi_{1 2} \bigl(- \tst{\frac{\be}{N}}\bigr)
     \Pi_{3 4} \bigl(- \tst{\frac{\be}{N}}\bigr) \dots
     \Pi_{N-1 N} \bigl(- \tst{\frac{\be}{N}}\bigr) =
        \tau^\frac{N}{2} \bigl(- \tst{\frac{\be}{N}}\bigr) \epp
\end{equation}
Inserting the latter equation into (\ref{surffree1}) we can formulate
the following theorem.
\begin{theorem} \label{main}
The surface free energy of the open XXZ chain is determined by
	\begin{equation} \label{surffree2}
	     \D F = \lim_{\substack{N \rightarrow \infty\\
	                  N \in 2{\mathbb N}}}
             - T \ln \frac{\<\{\la\}| \tau^\frac{N}{2}
	                      \bigl( - \tst{\frac{\be}{N}} \bigr)
			      |\{\la\}\>}{\<\{\la\}|\{\la\}\>} \epc
\end{equation}
where $|\{\la\}\>$ is the unnormalized dominant state of the rescaled
quantum transfer matrix $\tr  \tilde{T}^{QTM} (\la)$, equation
(\ref{rescqtm}), and $\tau$ is the finite temperature boundary
operator as defined in (\ref{deftau}).
\end{theorem}
Recall that within the scheme of the algebraic Bethe Ansatz the
vectors $|\{\la\}\>$ and $\<\{\la\}|$ are given by
\begin{equation}
     |\{\la\}\> = B(\la_1) \dots B(\la_\frac{N}{2}) |0\> \epc \qd 
     \<\{\la\}| = \<0| C(\la_\frac{N}{2}) \dots C(\la_1) \epc
\end{equation}
where $B(\la)$ and $C(\la)$ have been defined in (\ref{rescqtm}),
where $|0\> = \bigl[\bigl( \begin{smallmatrix} 1 \\ 0 \end{smallmatrix}
\bigr) \otimes \bigl( \begin{smallmatrix} 0 \\ 1 \end{smallmatrix}
\bigr)\bigr]^{\otimes N/2}$ is the staggered pseudo vacuum of the
quantum transfer matrix with dual $\<0|$, and where
$\{\la_j\}_{j=1}^{N/2}$ is the solution of the Bethe Ansatz equations
\begin{equation}
     \biggl[
       \frac{\sh(\la_j + \frac{\be}{N}) \sh(\la_j - \frac{\be}{N} + \h)}
            {\sh(\la_j - \frac{\be}{N}) \sh(\la_j + \frac{\be}{N} - \h)}
       \biggr]^\frac{N}{2}
     = \prod_{\substack{k=1 \\k \ne j}}^{N/2}
       \frac{\sh(\la_j - \la_k + \h)}{\sh(\la_j - \la_k - \h)} \epc
\end{equation}
$j = 1, \dots, N/2$, that maximizes the modulus of the leading
eigenvalue of the quantum transfer matrix at $\la = 0$.
\begin{example}
As an example let us again consider the XXZ chain with reflecting ends,
i.e.\ vanishing boundary fields. We have to set $\x^\pm = \i \p/2$ in
(\ref{kmother}), leading to $2 K^{(+)} (\la) = K^{(-)} (\la) = I_2$,
the $2 \times 2$ unit matrix. Inserting this into the definition
(\ref{deftau}) of the finite temperature boundary operator we obtain
\begin{align}
     \tau (\la) & = \tst{\2} \tr \tilde T^{QTM} (- \la)
                    \s^y \tilde{T}^{QTM \, t} (\la + \h) \s^y \\ \notag
                & = \tst{\2} \bigl( A(-\la) D(\la + \h)
		                  - B(-\la) C(\la + \h)
		                  - C(-\la) B(\la + \h)
		                  + D(-\la) A(\la + \h) \bigr) \epp
\end{align}
\end{example}

In the remaining part of this section we would like to demonstrate
that the finite temperature boundary operator $\tau (\la)$ is
generically connected with the reflection algebra (\ref{refl}).
Our basic result is
\begin{proposition} \label{prop:rrefl}
With any c-number representation $K^{(+)} (\la)$ of the right
reflection algebra (\ref{rrefl}) the matrix
\begin{equation} \label{ourreflrep}
     \CT^{(+) t} (\la) = \tilde T^{QTM} (- \la) \,
        K^{(+)} (\la) \, \s^y \tilde{T}^{QTM \, t} (\la + \h) \s^y
\end{equation}
as well defines a representation of (\ref{rrefl}).
\end{proposition}
\begin{proof}
We only sketch the proof. The idea is to show that
\begin{equation}
     T^{(+)} (\la) := \tilde{T}^{QTM \, t} (- \la) \epc
\end{equation}
which clearly defines a representation of the Yang-Baxter algebra
(\ref{yba}), satisfies (\ref{crossing}), and then to employ proposition
\ref{prop:comult}. In order to show that $T^{(+)} (\la)$ satisfies
(\ref{crossing}) one may, for instance, use (\ref{tqtmrtr}),
(\ref{rescqtm}) and apply to the elementary $L$-matrix (\ref{lmatrix})
equation (\ref{lcrossing}) and the general relation
\begin{equation} \label{invert}
     T^{-1} (\la) = \frac{\s^y T^t (\la - \h) \s^y}{\detq T(\la)}
\end{equation}
for the inverse of a monodromy matrix associated with the $R$-matrix
(\ref{rxxz}). Notice that $\detq T(\la)$ in (\ref{invert}) denotes the
quantum determinant (see e.g.\ \cite{Sklyanin88} or chapter 14 of
\cite{thebook}). For $T^{(+)} (\la)$ the function $\dh (\la)$ in
(\ref{crossing}) turns out to be
\begin{equation}
     \dh (\la) = \biggl[
        \frac{b(- \la + \frac{\be}{N} - \h) b(- \la - \frac{\be}{N})}
             {b(- \la - \frac{\be}{N} + \h) b(- \la + \frac{\be}{N})}
	     \biggr]^\frac{N}{2} \epp
\end{equation}
Knowing that $T^{(+)} (\la)$ satisfies (\ref{crossing}) we apply
proposition \ref{prop:comult} to the special case, when
$\tilde{\CT}^{(+)} (\la) = K^{(+)} (\la)$ is a c-number representation
of the right reflection algebra (\ref{rrefl}). Then also $K^{(+) \, t}
(\la)$ is a c-number representation of (\ref{rrefl}), which can be
seen by taking the transpose of (\ref{rrefl}) and using the symmetry
of the $R$-matrix. Taking into account that
\begin{equation}
     T^{(+) \, a} (- \la)
        = \frac{\tilde{T}^{QTM \, -1} (\la + 2 \h)}{\dh(- \la)}
        = \frac{ \s^y \tilde{T}^{QTM \, t} (\la + \h) \s^y}
	       {\dh(- \la) \detq \tilde{T}^{QTM} (\la + 2 \h)}
	       \epc
\end{equation}
the proposition follows.
\end{proof}
Proposition \ref{prop:rrefl} has a number of simple but important
consequences.
\begin{corollary} \label{cor:taufea}
For any pair of c-number solutions $K^{(\pm)} (\la)$ of the reflection
equations (\ref{refl}) the following holds.
\begin{enumerate}
\item
The finite temperature boundary operator $\tau(\la)$, equation
(\ref{deftau}), defines a commuting family of operators,
\begin{equation}
     [\tau(\la), \tau(\m)] = 0 \qd
        \text{for all $\la, \m \in {\mathbb C}$} \epp
\end{equation}
\item
$\tau (\la)$ can be diagonalized by means of the algebraic Bethe
Ansatz.
\item
The matrix element
\begin{equation}
     \PH (\n_1, \dots, \n_\frac{N}{2})
        = \<\{\la\}| \prod_{j=1}^{N/2} \tau (\n_j) |\{\la\}\>
\end{equation}
is a symmetric function of the parameters $\n_j \in {\mathbb C}$,
$j = 1, \dots, N/2$.
\end{enumerate}
\end{corollary}

\section{Conclusions}
We have expressed the surface free energy for the XXZ chain in the
thermodynamic limit as Trotter limits (\ref{surffree1}) and
(\ref{surffree2}) of expectation values of certain operators in the
dominant eigenstate of the quantum transfer matrix.

In (\ref{surffree1}) the operator is local and is the product of
certain projection operators $\Pi$ defined in (\ref{defpi}). This
formula is not only valid for the XXZ chain with reflecting ends, but
was derived for a rather large class of open boundary systems,
specified by a pair of c-number solutions of the reflection equations
(\ref{refl}). If the corresponding $R$-matrix of `difference form' and
satisfying (\ref{lrsym})-(\ref{sym}) is regular we can always associate
a quantum transfer matrix with it, and the projection operators
(\ref{defpi}) are well-defined.

The operator in (\ref{surffree2}) is non-local and is a simple
quadratic expression in terms of the matrix elements of the rescaled
quantum monodromy matrix (\ref{rescqtm}). We called it the finite
temperature boundary operator. In proposition \ref{prop:rrefl} and
corollary \ref{cor:taufea} we worked out its relation to the
reflection algebra (\ref{refl}). We derived (\ref{surffree2})
for the case of the XXZ chain with boundary conditions specified
by any c-number representation of the reflection algebra (\ref{refl}).
Still, we used only a few rather general properties of monodromy
matrices connected to the $R$-matrix of the XXZ chain, namely the
crossing relation in the form (\ref{lcrossing}) and the existence of a
quantum determinant, which allowed us to apply the inversion formula
(\ref{invert}). It seems rather likely that a finite temperature
boundary operator can be defined for all solvable systems having
these additional properties.

The explicit calculation of the surface free energy from our formulae
remains a challenging task, since the calculation of matrix elements,
such as in (\ref{surffree1}) and (\ref{surffree2}), within the
algebraic Bethe Ansatz is, in general, involved. In the light of the
results obtained e.g.\ in \cite{KMT99b,GKS05} there is, however, hope
to accomplish this challenging task in the future.

{\bf Acknowledgement.}
The authors would like to thank A. Kl\"umper and J. Sirker for
interesting discussions which triggered and stimulated this work.
MB gratefully acknowledges financial support by the German Science
Foundation under grant number Bo 2538/1-1.


\providecommand{\bysame}{\leavevmode\hbox to3em{\hrulefill}\thinspace}
\providecommand{\MR}{\relax\ifhmode\unskip\space\fi MR }
\providecommand{\MRhref}[2]{%
  \href{http://www.ams.org/mathscinet-getitem?mr=#1}{#2}
}
\providecommand{\href}[2]{#2}

\end{document}